\documentclass[preprint2]{aastex}

\slugcomment{}

\shorttitle{3-D modeling and distance of Mz1}
\shortauthors{Monteiro et al.}

\begin{document}

\title{3-D Photoionization Structure and Distances of 
Planetary Nebulae II.  Menzel\,1}

\author{Hektor Monteiro$^{1,2}$, Hugo E. Schwarz$^1$, Ruth Gruenwald$^2$, 
Katherine Guenthner$^3$, Steve R. Heathcote$^1$}
\affil{1 Cerro Tololo Inter-American Observatory\altaffilmark{1}, Casilla 603, 
Colina El Pino S/N, La Serena, Chile}
\affil{2 Instituto de Astronomia, Geof\'isica e Ciencias Atmosf\'{e}ricas, 
S\~{a}o Paulo, Brasil}
\affil{3 Astronomy Dept., University of Texas at Austin, Austin, TX 78712, USA}

\altaffiltext{1}{Cerro Tololo Inter-American Observatory, National Optical 
Astronomy Observatory, operated by the Association of Universities for
Research in Astronomy, Inc., under a cooperative agreement with the
National Science Foundation.}
\altaffiltext{2}{CTIO REU-PIA Program student.}

\begin{abstract}
We present the results of a spatio-kinematic study of the planetary
nebula Menzel\,1 using spectro-photometric mapping and a 3-D
photoionization code. We create several 2-D emission line images from
our long-slit spectra, and use these to derive the line fluxes for 15
lines, the H$\alpha$/H$\beta$ extinction map, and the [SII] line ratio
density map of the nebula. We use our photoionization code constrained
by these data to derive the three-dimensional nebular structure and ionizing
star parameters of Menzel\,1 by simultaneously fitting the
integrated line intensities, the density map, and the observed
morphologies in several lines, as well as the velocity
structure. Using theoretical evolutionary tracks of intermediate and
low mass stars, we derive a mass for the central star of
0.63\,$\pm$0.05M$_{\odot}$. We also derive a distance of 1050\,$\pm$150pc to Menzel\,1.
\end{abstract}

\keywords{Planetary Nebula -- Interstellar medium}

\section{Introduction}

The Planetary Nebula (PN) Menzel\,1 (Mz\,1) or G322.4-02.6 (15$^h$
34$^m$ 16$^s$.7 -59$^o$ 08' 59'' 2000.0) is a bright object with a bipolar
morphology and a prominent central ring of enhanced emission. H$\alpha$ and
[OIII] narrow band images of Mz\,1 have been published by
\cite{SCM92}. Being bright has not resulted in Mz\,1 being
well-studied; only a few papers have been dedicated to the object.

H$_2$ emission has been detected from Mz\,1 by \cite{W88} and the
morphology in this molecular line is similar to that of our optical
image, shown in Fig.\,\ref{pic}.

A detailed kinematical study of Mz\,1 was done by \cite{MB98} using
high resolution echelle spectroscopy. Using the intensity ratio of the
{[}SII{]} lines they computed densities of 1700 $cm^{-3}$ for the main
ring and 400 $cm^{-3}$ for the bipolar part. They show that the gas in
the main part of the nebula is expanding at 23\,$km \, s^{-1}$, and
that the velocity structure is consistent with a cylindrical model
with expansion velocities proportional to the radial distance from the
center. They also show that the dynamical age of the ring in Mz1 is of
the order of 7000\,yrs and estimate a mass of about 0.5M$_{\odot}$ for
the nebula using dynamical arguments and their computed distance of 2.0\,kpc.

Distances have been determined by several authors such as
\cite{VZ95} with 2.53 kpc, \cite{CKS92} with 2.28 kpc 
and \cite{Z95} with 2.85 kpc, all using statistical methods and based
on the same data. Using the same formalism as \cite{VZ95} and a new
value for the radius of the nebula, Marston et al. (1998) obtained a
distance of 2.0 kpc with a claimed uncertainty in this estimate of
about 30\%. \cite{AO92} lists 10 distances of which 8 are statistical
and 2 individual determinations. Given that all the above authors use
a filling factor, $\epsilon$\,=\,0.5, any such method has an inherent
and large uncertainty in the distances they determine. We compute
2.0\,$\pm$0.5\,kpc (adjusted standard deviation) from all 14
literature distances without any weighting factor. Note that each
individual distance can be in error by a large factor (see the
appendix), and the standard deviation computed from the literature
values is not very robust.

An effective temperature of 139\,kK and a luminosity of
147\,L$_{\odot}$ were determined by \cite{SCS93} with a distance of
1.8\,kpc. Mz\,1 is clearly under-luminous for a typical PN.

In this work we present observations of, and a 3-D photoionization
model for Mz\,1, and derive the 3-D structure of the PN constrained by
observed fluxes and morphologies in many emission lines, using the
same method as \cite{MSGH04} applied to NGC\,6369. By determining
the 3-D structure of nebulae, the large uncertainty involved in all
classical statistical distance determination methods is
eliminated. Assuming an arbitrary filling factor, constant ionized
mass or diameter, mass-radius relationship etc. is not needed here: we
{\it determine} what the structure and ionized mass are, and can
therefore derive distances to much greater accuracy than has been
previously possible.

In summary, we obtain the 3-D spatial structure of the nebula along
with its chemical composition, ionizing source temperature,
luminosity, and mass, as well as an {\it independent} distance, in a
self-consistent manner. In \S2 we discuss the observations and briefly
explain the basic reduction procedures, including our image
reconstruction technique used to obtain the emission line intensity
images. In \S3 the results obtained from these images are discussed:
the pixel by pixel reddening correction of the images, the integrated
line fluxes, and the computed temperature and density maps. In \S4 we
present the model results generated by our 3-D photoionization code,
and we discuss the derived quantities. In \S5 we give our conclusions,
and explain in detail how our method works in the appendix.

\section{Observations}

\subsection{Observations and data reduction}

\begin{figure}
\includegraphics[width=\columnwidth]{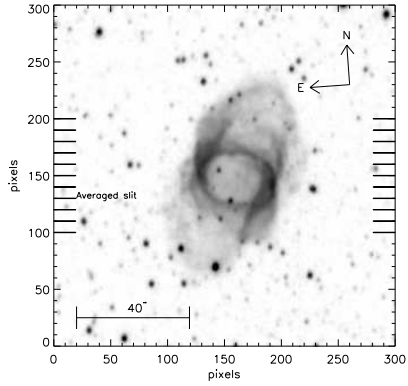}
\caption{Our narrowband {[}SII{]} image of Mz\,1 with the scale 
and orientation indicated. We have indicated the centers of the slit 
positions we used by the short strokes on both sides of the image. \label{pic}}
\end{figure}

We show our image of Mz\,1 in Fig.\,\ref{pic} taken in the light of the
[SII]671.7nm line through a filter with $\lambda_c$\,=\,671.8nm and
FWHM\,=\,2.6\,nm.  The 300\,s exposure was taken with the CCD camera
attached to the CTIO 0.9m telescope on the 9th of April 2002. The
plate-scale is 0.4\arcsec/pix on the 2kx2k TEK chip, and the seeing
during the exposure was ~1.1\arcsec~according to the nearby seeing
monitor. 

\clearpage    \clearpage

The main features of this [SII] image are similar to those found in
the H$\alpha$ image of \cite{SCM92}. The morphology of Mz\,1 is
complex: two faint outer lobes toward the NW and SE meet in the
brighter central annular region, and there are enhanced extended bands
of emission on the eastern and western side of the central
annulus. Some fine structure is also seen throughout the surface of
the nebular emission.  

The spectra were taken with the CTIO 1.5\,m Ritchey-Chr\'etien
telescope with the RC Spectrograph on the nights of 13 and 14th June,
2002. We used a grating with 600\,l/mm blazed at 600\,nm giving a
spectral resolution of 0.65\,nm/pixel and a plate scale of
1.3\arcsec/pixel with a slit width of 4\arcsec.  The spectral coverage
obtained with this configuration was approximately 450\,nm to
700\,nm. For details of the instrument and telescope see
http://www.ctio.noao.edu and click on ``Optical Spectrographs'', then
on ``1.5m RC spectrograph''.

By taking exposures at several parallel long-slit positions across the
nebula, we obtained line intensity profiles for each slit. These
profiles were then combined to create emission line images of the
nebula with a spatial resolution of about $4\arcsec\times 4\arcsec$,
in a way similar to radio mapping. Due to a technical problem the
data from one of the observed slit positions had to be discarded. We
computed an average of the two adjacent slit spectra as representing
this position (shown in Fig.\,\ref{pic}). The added uncertainties
introduced by this procedure are discussed below.

The individual slits were observed and reduced using standard
procedures for long-slit spectroscopy, using IRAF reduction packages.

A fine correction for slit misalignment was made using the H$\alpha $
and H$\beta $ profiles for each exposure. Using IDL, the images were
re-dimensioned to 100 times their original size. The normalized H$\alpha
$ and H$\beta $ profiles were then matched and the final result
re-dimensioned to original values. This procedure yields the precise
alignment necessary for calculation of diagnostic line intensity
ratios. Minor shifts of the order of one pixel can introduce
considerable errors in line ratios, if this method is not applied.

It is also important to note that the 11 slit positions observed do
not fully cover the nebula, leaving out the faintest outer parts of the
bipolar lobes; see Fig.\,\ref{pic}. Below we estimate this lost flux and show
that it is a small fraction of the total flux. Note, however, that in
our calculations, we take this ``lost flux'' into account by matching
our model output fluxes to the observed area, but the model does
provide the complete nebular structure.

\section{Observational results}

\begin{figure*}
\includegraphics[]{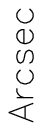}
\caption{Reddening corrected emission line images for Mz\,1. 
Only the 10 images with the highest S/N are shown.\label{maps_full} }
\end{figure*}

Images were created for all 15 lines detected with a 
signal-to-noise ratio above 5. In Fig.\,\ref{maps_full} the
constructed images for the most important emission lines are
shown. These images have been corrected for reddening as described in
the following section. The corresponding signal-to-noise images were
used to obtain the total line intensities with fractional errors,
given in Table 1.

\clearpage  

\subsection{Reddening correction and total fluxes}

\begin{figure}
\includegraphics[width=\columnwidth]{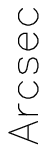}
\caption{The Mz\,1 H$\alpha $/H$\beta $ ratio map with contour 
overlay of H$\alpha$, where the darker shading represents stronger emission. 
Only the brighter parts of the nebula are 
shown as the image is cut for S/N values lower than 10.\label{ha/hb}}
\end{figure}

From our long-slit spectra we reconstruct emission line images using
the method described by \cite{MSGH04}.  These images for
each line were corrected for reddening using the H$\alpha $/H$\beta $
ratio map shown in Fig.\,\ref{ha/hb}.  The logarithmic correction
constant was calculated pixel by pixel using the theoretical value of
H$\alpha $/H$\beta $=2.87 from
\cite{OS89} and the reddening curve of \cite{S79}.

We investigated the effect of differential atmospheric refraction on
this ratio map. From the airmasses of our observed positions and the
values given by \cite{F82} we computed a correction which we applied
to our data. All slit positions were observed at airmasses below 1.5
except one outer position in a faint part of the nebula, which had an
airmass of 1.8. Since we used a 4\arcsec wide slit and the object is
extended, the effect was small, but not negligible in the steep
gradients near the bright ring structure. The average error due to
this effect is about 3\% in the high signal to noise (S/N) areas and
about 25\% in the low S/N areas. The nett effect on the final
calculated relative total fluxes is about 0.3\% for strong lines and
7\% for weak ones, well within the other observed uncertainties.

\clearpage    \clearpage

The final calculated line fluxes relative to H$\beta $ and their
corresponding 1$\sigma $ errors are shown in Table\,\ref{line
fluxes}. All total fluxes were obtained by integrating the reddening
corrected images pixel by pixel. The value we obtained for the total
reddening corrected H$\beta$ flux is $F_{H\beta}\,=\,2.6\times
10^{-11}erg\,cm^{-2}\, s^{-1}$. We estimate from the relative area of
the nebula that falls outside the slit images, and the fact that for
several lines the flux comes mainly from the central, bright part of
Mz\,1, that the lost flux in $H\beta$ is about 3\% of the total, and
the average lost flux in all lines about 5\%. Note that \cite{AO92}
lists an {\it uncorrected} H$\beta$ flux of 4.9$\times
10^{-12}erg\,cm^{-2}\, s^{-1}$, taken from \cite{P71}, which is close
to our uncorrected value of 5.8$\times 10^{-12}erg\,cm^{-2}\,
s^{-1}$. Our flux is obtained digitally by integrating our
spectrophotometry pixel by pixel over the whole nebula, while Perek
used photoelectric aperture photometry during non-photometric nights,
with all its associated calibration problems, and warns us in his
article that cirrus affected many of the measurements, explaining why
his flux is lower than ours which was taken during a photometric
night. Perek also listed a previous measure of the H$\beta$ flux of
-11.26 which is equal to our uncorrected flux.

\begin{table}

\centering

\caption{ Line fluxes relative to H$\beta $\label{line fluxes}}

\begin{tabular}{lccc} \hline {Line $\lambda$(nm)}& {Flux}& {Dered. Flux}& 
{Error\,(\%)}\\ 
\hline\hline

     {HeII$\lambda$468.6} & {0.11} & {0.11} & {7}\\

     {[OIII]$\lambda$495.9} & {2.35} & {2.28} & {6}\\

     {[OIII]$\lambda$500.7} & {7.11} & {6.83} & {6}\\

     {[NII]$\lambda$575.5} & {0.10} & {0.07} & {23}\\

     {HeI$\lambda$587.6} & {0.25} & {0.16} & {9}\\

     {[OI]$\lambda$630.0} & {0.33} & {0.21} & {8}\\

     {[SIII]$\lambda$631.1}& {0.03} & {0.02} & {20}\\

     {[OI]$\lambda$636.3} & {0.12} & {0.08} & {14}\\

     {[NII]$\lambda$654.9} & {2.96} & {1.81} & {12}\\

     {$H\alpha$} & {4.68} & {2.88} & {6}\\

     {[NII]$\lambda$658.4} & {9.18} & {5.57} & {12}\\

     {HeI$\lambda$667.8} & {0.08} & {0.05} & {15}\\

     {[SII]$\lambda$671.7} & {0.82} & {0.49} & {12}\\

     {[SII]$\lambda$673.1} & {0.76} & {0.45} & {12}\\ 

     \hline

\end{tabular}
\end{table}

Using these images we estimated the effect of adopting the average for
the discarded slit position. Since the nebula shows considerable
symmetry, we compared the northern half of the reconstructed images
with the southern half. By comparing the total fluxes from these
halves we determined that the error on the calculated total fluxes
were about 10\% for low ionization lines and about 2\% for the other
ones. This difference is due to the fact that the discarded position
is close to the bright ring of the nebula, where the low ionization
lines are relatively strong. The effect is relatively small and this
added uncertainty is taken into account in the values presented in
Table 1.

\subsection{Gas density and temperature}

\begin{figure}
\includegraphics[width=\columnwidth]{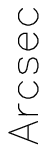}
\caption{Density map obtained from the {[}SII{]}671.7,673.1nm line 
ratio.\label{dens_map}}
\end{figure}

We calculated density and temperature maps from the reddening
corrected maps of the {[}SII{]} and {[}NII{]} lines respectively. The
expressions relating the line intensity ratios to the gas density and
temperature are the ones published by \cite{M84} and those used in the
IRAF {\it temden} package. Details are as in \cite{MSGH04}. The
density map is shown in Fig.\,\ref{dens_map}.

\begin{figure}[h]
\includegraphics[width=\columnwidth]{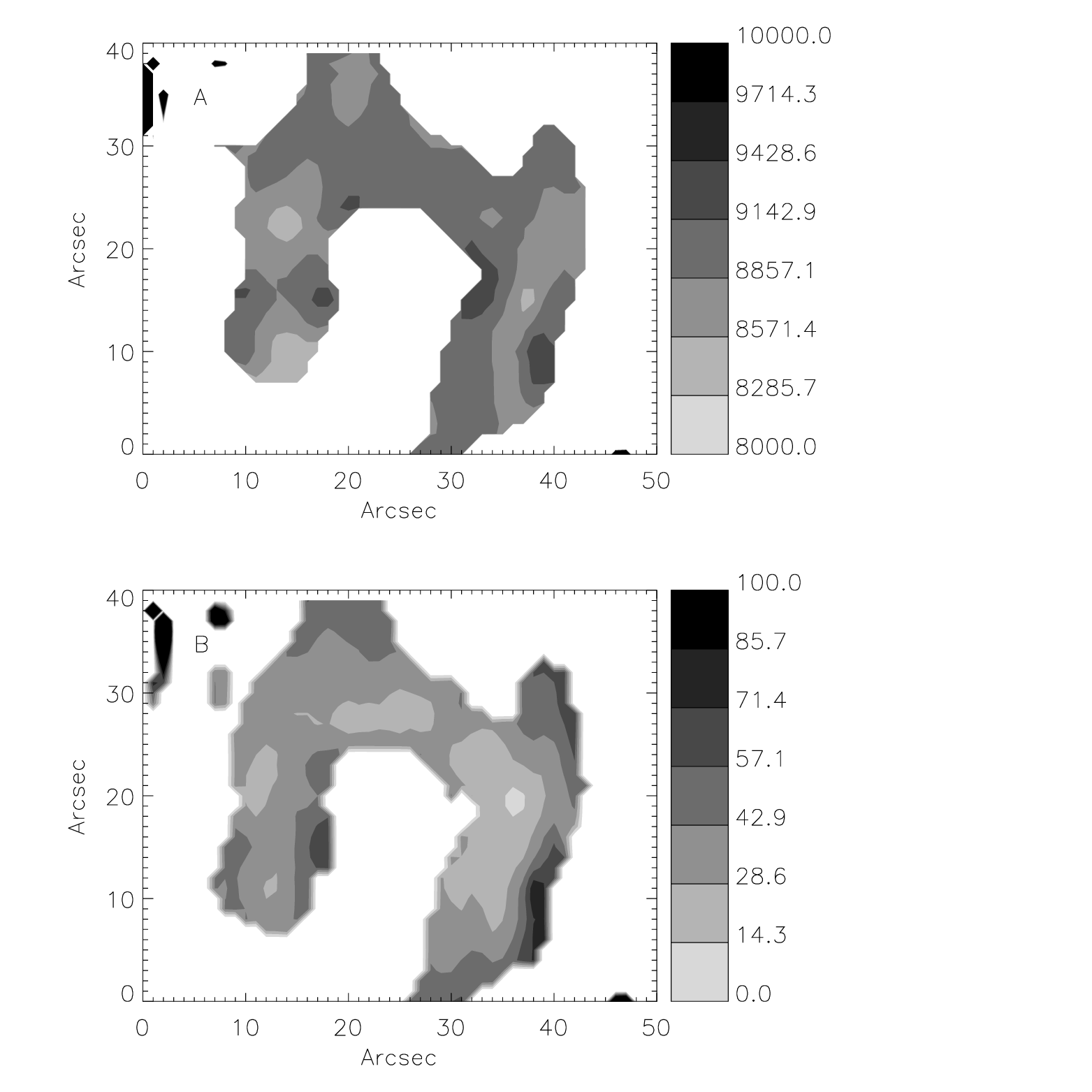}
\caption{(A) Temperature map obtained from the de-reddened {[}N II{]} 
line ratio. The scale on the right is in degrees centigrade; 
(B) difference between this map and the one obtained 
assuming constant density. Note that the maximum difference is 
less than 100\,K. North is up, and east is to the left.\label{temp_comp}}
\end{figure}

For the temperature maps, the correction for slit misalignment was
carried out as done for the H$\alpha $/H$\beta $ maps discussed in
section 2.1. We compute the temperature map for Mz\,1 in two ways:
using our density map we obtain the temperature map shown in
Fig.\,\ref{temp_comp} (A) and in (B) we show the difference between
this map and the one calculated assuming constant density. The
maximum difference between these two maps is less than 100\,K. The
images are clipped for data values with S/N lower than 10 for visual
clarity.

\clearpage    \clearpage

\clearpage    \clearpage

\section{Photoionization Models}

The photoionization code applied here has been described in detail by
\cite{GVB97}. It uses a cube divided into cells, each
having a given density, allowing arbitrary density distributions to
be studied. Typical models runs use cubes of 70 to 100 cells on a
side. The input parameters are the ionization source parameters
(luminosity, spectrum, and temperature), elemental abundances, density
distribution, and the distance to the object. The conditions are
assumed uniform within each cell for which the code calculates the
temperature and ionic fractions. These values are used to obtain
emission line emissivities for each cell.

The final data cube can be spatially oriented and projected in order
to reproduce the observed morphology. The orientation on the sky of
the 3--D nebular structure is thus determined. The line intensities
and other relevant quantities are then obtained after projection onto
the line of sight.

\begin{figure}[h]
\includegraphics[width=\columnwidth]{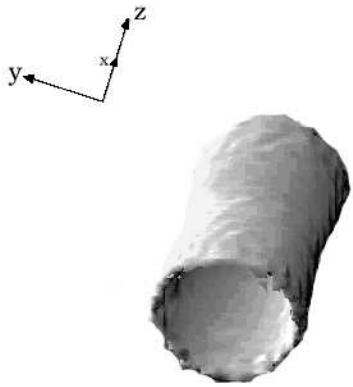}
\caption{Isocontour plot of the 3-D density structure determined for Mz\,1 
using model calculations. The inclination angle of the polar axis 
of the structure is 40$^o$.\label{strut3D}}
\end{figure}

\begin{figure}[h]
\includegraphics[width=\columnwidth]{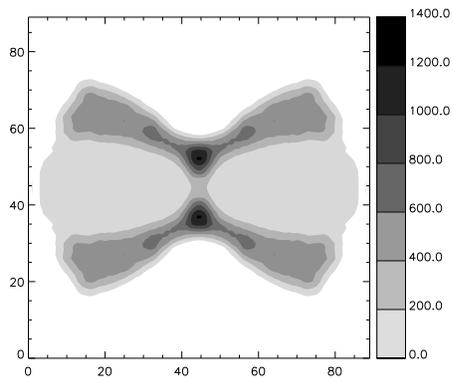}
\caption{Cut of the density structure of Mz\,1 determined 
by the model calculations.\label{strut3D_cut}}
\end{figure}

The structure we obtained for Mz\,1 is shown in Fig.\,\ref{strut3D} as
it is oriented relative to the observer. We also show a cut along the
major axis of symmetry in Fig.\,\ref{strut3D_cut}, indicating the
density values. The 3--D structure of Mz\,1 was constrained by our
density map to be an open structure, not a closed shell of any type.
We therefore adopt an hour-glass shape with a density gradient from the equator
to pole. We added random density fluctuations to better fit the line
fluxes, especially H$\beta$. The rotation angles relative to the x, y,
and z axes respectively, are 0$^o$,10$^o$, and 40$^o$, with the
symmetry axis of the main hour-glass structure being x. The model
resolution of $100^3$ cells was limited by our 4\,GB of computer
memory, and code execution time.

\clearpage    \clearpage

\clearpage    \clearpage

The ionizing spectrum we used for the central star (CS) was a blackbody
modified by the He and H atomic absorption edges at 54.4eV and
13.6eV. The addition of these edges was necessary to be able fit the
{[}OIII{]} and HeII line intensities simultaneously, the HeII line
being particularly sensitive to the depth of the absorption edge, and
therefore providing a good central star temperature constraint. These
absorption edges are also present in the more sophisticated
theoretical atmospheric models presented by \cite{RDDW00}, which
are similar to the ones we used here for similar stars. The only
feature from Rauch's spectra that we do not incorporate is the added
absorption due to other lines, but this has no significant
effect on our model. The effective temperature and luminosity of our
adopted crude stellar spectrum are given in Table\,\ref{mod-res}.

\subsection{Model Results}

\begin{figure*}
\includegraphics[]{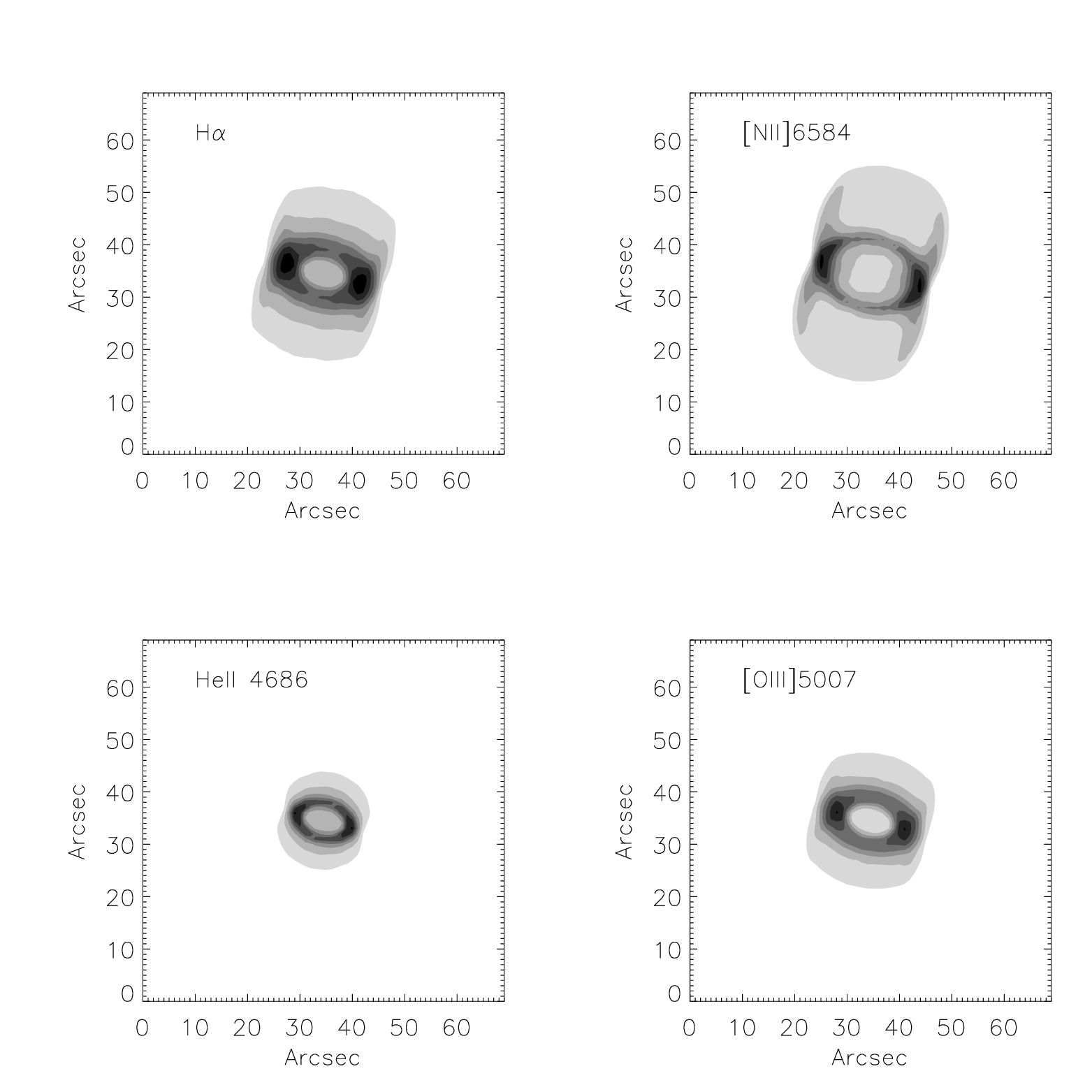}
\caption{Projected line images obtained with the photoionization 
model. \label{ims-mod}}
\end{figure*}

We present here the main results obtained from the photoionization
model constrained by the observational data. The total line
intensities are given in Table\,\ref{mod-res}, as well as the fitted
abundances and ionizing star parameters. Projected line images
for four important transitions are shown in Fig.\,\ref{ims-mod}.

\begin{figure}[h]
\includegraphics[width=\columnwidth]{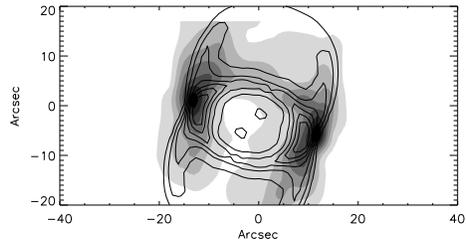}
\caption{Contour image of model {[}NII{]} image overplotted 
on observed image for our best distance of d=1050$\pm$150\,pc.
\label{mod-comp}}
\end{figure}

The model image size is fitted to the observed one for the line
{[}NII{]}658.4nm, as well as the absolute $H\beta$ flux, giving a
final distance of 1050$\pm$150\,pc. Fig.\,\ref{mod-comp} shows the
observed {[}NII{]} image with the corresponding model image contours
overlaid for the obtained distance.
   
\clearpage  

\clearpage    \clearpage

\begin{table}
\begin{center}
\caption{Observed and model line fluxes and model central star 
parameters.\label{mod-res}}
\begin{tabular}{lll}\hline  &   {Observed}  &   {Model}\\
\hline\hline 

{$T_*$ (K)} & {} & {120kK}\\
{$L_*/L_{\odot}$} & {-} & {164}\\
{Density} & {100-1400} & {100-1400}\\
{He/H}      & {-}   & {$1.14\times 10^{-1}$}\\
{C/H}       & {-}   & {$3.3\times 10^{-4}$}\\
{N/H}       & {-}   & {$2.2\times 10^{-4}$}\\
{O/H}       & {-}   & {$4.7\times 10^{-4}$}\\
{Ne/H}      & {-}   & {$3.5\times 10^{-4}$}\\
{S/H}       & {-}   & {$1.1\times 10^{-5}$}\\
{log($H\beta$)}  &  {-10.6}  &  {-10.6}\\
{[NeIII]386.8$^{a}$}&  {5.64}& {4.8}\\
{HeII468.6} &   {0.014}  &   {0.013}\\
{[OIII]500.7}&   {6.83} &  {6.68}\\
{HeI587.6}  &   {0.16}     &   {0.17}\\
{[OI]630.0} &   {0.21}     &   {0.26}\\
{[NII]658.4}  &  {5.6}  &  {5.3}\\
{[SII]671.7}&   {0.49}     & {0.49}\\
{[SII]673.1}&   {0.45}     & {0.46}\\     

\hline

\end{tabular}
\end{center}
a) Value obtained by \cite{KSB97}
\end{table}

\begin{figure}
\includegraphics[width=\columnwidth]{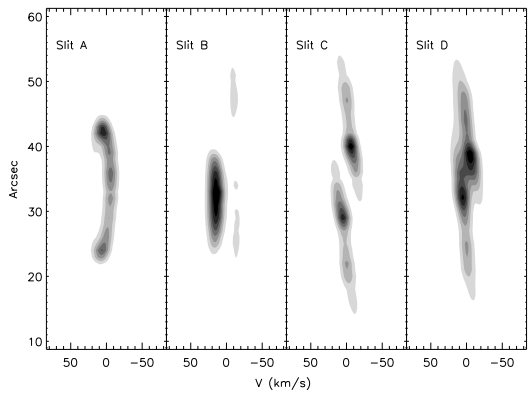}
\caption{Model PV diagram obtained for the four slit positions as observed 
by \cite{MB98} in the light of the  [NII]658.4\,nm line.\label{PVmz1}}
\end{figure}

Using the model cube of temperatures and ionization structure we also
calculated the position-velocity (PV) diagram for four different slit
positions using a spherically symmetrical velocity field given by
$\vec V = \alpha (\left|\vec r\right|/rmax)\times \vec r/{\left|\vec
r\right|}$ . r$_{max}$ is half the side of our model data cube (here
we used r$_{max}$\,=\,8\,10$^{17}$\,cm), and $\alpha$ is the maximum
velocity reached within the data cube (here we used 45\,km/s). We
computed the PV diagram for the {[}NII{]}658.4nm line in the same
positions as those observed by Marston et al.(1998). The projected
velocities in the $x$ direction are obtained using:

$$ \phi_\lambda (v,y,z) = \sum_x {{\epsilon_\lambda (x,y,z)} \over
{\sqrt{\pi}. \xi(x,y,z)}} . e^{ - {[{{\Delta V(v,x,y,z)} \over
{\xi(x,y,z)}}]}^2}~~~~~~~~~~~~~~~~(1)$$

\noindent 

with: $$ \Delta V(v,x,y,z) = V_x(x,y,z) - v~~~~~~~~~~~~~~~~~~~~~~~~~~~~~~~~(2)$$

$$ \xi(x,y,z) = \sqrt{V_{th}^2(x,y,z) + V_T^2 }~~~~~~~~~~~~~~~~~~~~~~~~~~~~~~~~(3)$$

 $$ V_{th}(x,y,z) = \sqrt{2kT_e(x,y,z)/Am_H}~~~~~~~~~~~~~~~~~~~~~~~~~~~~(4)$$

\noindent 

where $\epsilon_\lambda (x,y,z)$ is the emissivity of a cell at
$(x,y,z)$, $V_T$ is the turbulent velocity (taken to be 2\,km/s) and
$V_{th}(x,y,z)$ the thermal velocity of an atom of atomic mass A, the
local electron temperature is $T_e(x,y,z)$, and v is the plotting
interval for the velocity, where in this case v goes from -80 to
+80\,km/s. m$_H$ is the mass of an H atom.  The PV diagram obtained is
shown in Fig.\,\ref{PVmz1}.

\clearpage    \clearpage

Comparing our model derived PV diagrams with the observed ones in
Figure~3 of Marston et al.(1998) we see a remarkable agreement. The
main features are reproduced, confirming our confidence in the 3-D
structure we adopted for Mz\,1.

\section{Discussion and conclusions}

We present spectrophotometric maps of Mz\,1, giving spatially
resolved information for many emission lines and precise total fluxes
for the observed part ($\ge$95\% of the total flux) of the nebula. The
images produced with this technique were used to determine properties
of the nebula using the H$\alpha $/H$\beta$ ratio map for the
de-reddening, the {[}SII{]} line ratios for the density, and the
{[{NII{]} line map for the temperature.

The H$\alpha $/H$\beta$ map shows little structure. The prominent ring
feature of the nebula does not show significant differences in
extinction when compared to the other regions.  This may imply that
most of the reddening is due to the foreground, and not intrinsic to
the nebula.

We also show the density map obtained from the observations indicating
the presence of a dense waist ring and a bipolar structure of lower
density. Based on this map we propose for Mz\,1 a 3--D hour-glass
structure with a waist whose density decreases smoothly from the
equator to the poles.

Using a photoionization code and the proposed structure we obtained a
complete 3--D physical model for Mz\,1. The fitted model line
intensities show excellent agreement --well within the computed
errors-- with the observed values. The obtained distance of
d=1050$\pm$150\,pc falls near the lower end of the error range
determined from 14 literature values for the distance. This makes
sense as the angular radius of 25\arcsec~used in the statistical
methods is smaller than the true size of the nebula of 38\arcsec~
as measured by \cite{SCM92}, and our distance is also smaller.

The luminosity of 164\,$\pm$\,25\,L$_{\odot}$ and temperature of
120\,$\pm$\,16\,kK we have determined for the central star are within
their errors equal to those found by \cite{SCS93}, taking into account
the difference in distance and total flux. Our errors have been
conservatively estimated to be similar to the cumulative observational
errors we computed and used to constrain the model. This luminosity is
low for a PN CS, and indicates that the star is evolved and well down
its Sch\"{o}nberner track, as confirmed by its position in
Fig.\,\ref{evolMZ1}.

\clearpage  \begin{figure}[h]
\includegraphics[width=\columnwidth]{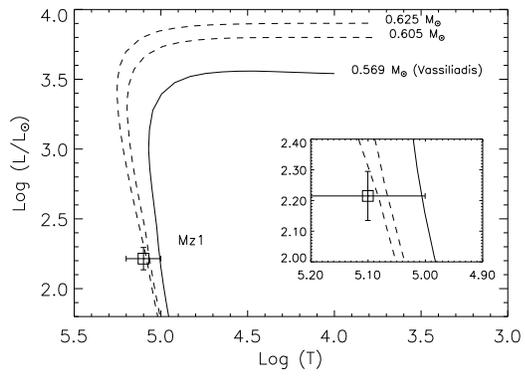}
\caption{Comparison of our model temperature and luminosity obtained for the 
central star with model tracks calculated by \cite{B95}. \label{evolMZ1}}
\end{figure}  \clearpage 

We determine the mass of the CS to be 0.63\,$\pm$0.05M$_{\odot}$ by
fitting to the theoretical evolutionary tracks of \cite{B95} , using
the errors estimated from our observational data and propagated to the
CS luminosity and temperature.  Due to the fact that in this part of
the HR diagram, the tracks lie close together, we cannot say with any
precision what the precursor mass of this star was, except to place a
lower limit on this mass of about 1\,M$_{\odot}$ using the added track from
\cite{VW94} in Fig.\,\ref{evolMZ1}. The likely value for the CS
precursor mass is about 3M$_{\odot}$, based on the most probable CS
mass of 0.63\,M$_{\odot}$. From our model output we compute the
ionized nebular mass to be 0.14\,$\pm$0.03\,M$_{\odot}$.

An independent check is the time scale of expansion of Mz1. From
\cite{MB98} and scaling to our distance we get an expansion time of
3500\,yrs for the ring which can be longer or shorter depending on
what the history of the expansion has been -decreasing velocity due to
energy conservation in a wind blown cavity would shorten the time and
increasing velocity due to expansion into a medium with decreasing
density would lengthen it-so a range of about 2000-5000\,yrs.  For the
outer parts of Mz1 this time would be about 4700-12000\,yrs as the
material has had to travel 2.4 times further away from the CS
(38\arcsec~instead of 16\arcsec). From our luminosity range and
computing Bl\"{o}cker track ages we obtain 4500 to 10000\,yrs, quite
compatible with the outer nebular expansion time, thus confirming our
distance determination.

Since the sum of the ionized nebular mass and the CS mass is
0.77\,M$_{\odot}$, we expect that there possibly is more than
2\,M$_{\odot}$ of neutral matter near Mz\,1. Clearly, for the lower CS
precursor mass limit of 1\,M$_{\odot}$, the neutral mass estimate is a
factor of ten smaller. In any case, there is likely more neutral than
ionized mass in the system. Note, however, that \cite{HBCF96} estimate
a molecular mass of 0.0086\,M$_{\odot}$ and an ionized mass of
0.027\,M$_{\odot}$.  They used a radius for Mz1 of 12.6\arcsec~instead
of the 38\arcsec~optical radius, so naively assuming that the mass
scales as the third power of the radius would increase their ionized
mass by a factor of up to 9 to 0.25\,M$_{\odot}$, which is much closer
to our computed mass of 0.14\,M$_{\odot}$.

We have put together the Spectral Energy Distribution (SED) from 0.36
to 100\,$\mu$m for Mz\,1, using data from the literature listed in
SIMBAD. The plot of $\lambda\,F(\lambda)$ is shown in
Fig.\,\ref{sed}. Note that the 100\,$\mu$m IRAS point is an upper
limit. The double peaked distribution is typical for a PN.

\clearpage  \begin{figure}
\includegraphics[width=\columnwidth]{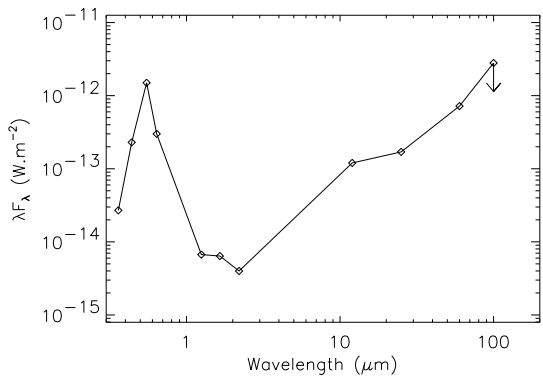}
\caption{The SED for Mz\,1 between 0.44\,$\mu$m and 100\,$\mu$m using flux 
values taken from the literature.\label{sed}}
\end{figure}  \clearpage

Integrating our SED yields a luminosity of about
60\,L$_{\odot}$ which includes the usual correction proposed by
\cite{M87} of a factor of 1.5. This implies that about 60\% of the UV
radiation from the CS escapes from the nebula. Given that the
structure we found is open toward the poles, and is likely clumpy,
this is a reasonable value. The fact that we estimate that there is
between 0.2 and 2\,M$_{\odot}$ of neutral matter around Mz\,1, some of
this may have, in fact, become ionized due to this escaping
radiation. Searching for a large, low surface brightness halo around
Mz\,1 may therefore be useful.

Perhaps most important is the fact that {\it our simple method can be
applied to any spatially resolved emission line nebula, yielding the
complete 3--D structure and an accurate distance}. The time consuming
multiple position long slit observations can be dispensed with since
Integral Field Spectrographs are now becoming available at major
observatories, making the method not just simple but also efficient.
See the detailed explanation of our method in the appendix.

\section{Appendix: Determining distances using 3-D photoionization models}

Classical distance determination methods are statistical or individual
in nature and all assume constancy of one or more physical parameters
of the PNe. \cite{G97} provides an excellent review of distance
determination methods and we refer to his book for details on other
methods than the ``Astrophysical Method'' (AM) which is the one we use
here.

The AM uses the fact that the electron density, H$\beta$ flux, and
angular extent are observable quantities of PNe --treated as
spherically symmetrical objects-- and that they are related to the
distance of the nebula by: \\

d\,=\,2.4.10$^{25}$\,F(H$\beta$)/[n$^2$\,.\,$\theta^3$\,.\,$\epsilon$]~~~~~~~~~~~~~~~~~~~~~~~~~~~~~~~~~~~~~~~~~~~~~~~~~~(5)

where F(H$\beta$) is the H$\beta$ flux in ergs/cm$^2$/s, n$^2$ is the
electron density in cm$^{-3}$, $\theta$ is the observed angular extent
of the nebula, and $\epsilon$ is the so-called filling factor which is
the fraction of the nebular volume which is emitting i.e. it contains
ionized gas.

It is therefore in principle possible, if all the above parameters are
known, to determine an accurate distance to any resolved, spherical
nebula or HII region.  In practice, however, the method has severe
limitations due to the necessity of restrictive assumptions such as
spherical symmetry. Taking each parameter in turn, we show that a
typical distance determination has an error of about a factor of 3 or
more for any given nebula. There are nebulae for which individual
distances with a range of a factor of 100 have been published (M2-9
extremes are: 50 to 5200\,pc;
\cite{S97})!

H$\beta$ fluxes are typically measured in a small aperture centered on
the PN, and an attempt is made to estimate the flux from the entire
nebula by extrapolation. Typical errors are large due to incomplete
knowledge of the size and brightness structure of the nebulae.  The
presence of the central star in the aperture increases the uncertainty
in this measurement.  Estimated errors are in the range 1.5 for
smooth, regularly shaped nebulae without much structure to 3 or more
for more typical PNe.

Electron densities, n, are usually determined from the ratio of the
pair of forbidden sulfur lines at 671.7 and 673.1nm. Since spectra
are also typically taken with small apertures centered on the PN or in the
best cases with a long slit but for only one position across the
nebula, again the uncertainty in the average value of n is large, say,
a factor of 1.5-2. Note that n appears in the equation to the second
power, increasing the effect of this error.

The angular extent, $\theta$, of a nebula can be measured but depends
on the passband or emission line used to make the observations. A PN
can have a diameter in the light of the [NII] line that is twice that
measured in H$\beta$ or [OIII], again producing uncertainties in the
distance determination. Smaller nebulae have a correspondingly larger
uncertainty associated with this measurement. $\theta$ appears in the
equation to the third power, increasing the effect of the uncertainty
in this parameter. An error of about a factor of 1.5 - 3 is again
typical.  In any case, the use of the radius assumes spherical
symmetry, which for most PNe is far from realistic.

The most difficult to determine and therefore least known parameter is
the filling factor, $\epsilon$. Usually, it is taken to be 0.5 in
statistical methods using large samples of nebulae, but the true
value can be anywhere between 0 and 1, with typical estimates varying
between 0.2 and 0.9, i.e. a factor of about 5 uncertainty.

Propagating all these factors, it is clear that determining a distance
to a PN is a very uncertain business.

Our method differs fundamentally from the classical methods, in which
one or more of the above parameters were assumed to be constant or
have some simple relationship. The use of a 3-D structure
to model the nebula eliminates the need for assumptions on
the quantities $n$,$\theta$, among others, as the structure can be
modified to include small scale variations in density such as clumps,
filaments and others as well as large scale variations (hour-glass
shapes for example). More importantly, this procedure removes entirely
the need to specify a filling factor $\epsilon$, as the large and
small scale density variations are all well defined in the 3-D
structure. We also determine all of the observables with high
precision from either long slit or Integral Field Unit (IFU) spectra
across the nebula. Our photoionization model is then constrained by
the quantities derived from these detailed spectra, and spectral
images.  We constrain simultaneously with: several line images,
several line fluxes, complete projected density map, and the velocity
structure, obtaining the best overall fit by adjusting the central
star luminosity, spectral distribution, temperature, average chemical
abundances, and the distance, also obtaining the complete and detailed
3-D structure of the nebula. All the above mentioned parameters are
therefore known much more precisely, and the distance determination is
correspondingly better. Typically we can compute the distance to about
10-20\% depending mainly on the observational errors.

\subsection{Model fitting procedure}

The details of the numerical photoionization code are given in the
appendix of \cite{GVB97}. In summary, the gaseous region, with the
radiation source in the center, is divided into cubic cells, for each
of which the physical conditions are, by definition, homogeneous. In
each cell thermal and ionization equilibrium is assumed in order to
obtain the physical conditions. The radiative transfer problem is
calculated with the ``on the spot'' approximation to save computing
time.

The input parameters are the elemental abundances, the gas density
distribution, the shape and intensity of the central ionizing
radiation spectrum (temperature, H and He absorption edges, and
luminosity in the case of a star) and the distance to the object. The
code then provides the physical conditions in each point of the
nebula, i.e., ionic fractional abundances, electronic temperature and
density, as well as the emission line luminosities of each cell.

This output is then used to produce projected images
and total fluxes for a given number of emission lines and a set of
(x,y,z) orientation angles. This output can then be tailored to match
given observational configurations such as single long slits, or
multiple slits (as is the case for Mz1). From the projected images we
can also construct projected diagnostic maps, such as density and
temperature maps.

The model generated or simulated ``observations'' discussed above are
then compared to the actual observational data obtained (in this case,
the spectro-photometric mapping of Mz1). We compare total line
intensities and check for discrepancies. If one or more of the
intensities are out of the range of the observational errors, we
proceed to fine tune input parameters that have influence on the given
line. For example, we take the $H\beta$ total fluxes from the model
and observations and compare them. The $H\beta$ line is mainly
dependent on the star luminosity and the 3-D structure of the gas, so
we adjust these input parameters accordingly. In this case it is
important to realize that the 3-D structure is actually defined by the
density in each cell and a physical size for the object which is
dependent on the distance. So in fact we are dealing with two input
parameters when we consider the $H\beta$ flux. The same type of
comparison is made for other lines such as [OIII]500.7nm which is an
important coolant, HeII468.6nm which depends mainly on the CS
temperature, among others. We also compare the model projected images
and diagnostic maps to those obtained from the observations.

After fitting all the model constraints to their respective
observational counterparts and adjusting the input parameters of the
code accordingly, we calculate a new model. The same procedure
discussed above is then repeated until we reach a satisfactory
agreement between model and observations for all line images, fluxes,
diagnostic maps, etc.

Notice that, after this iterative procedure, we obtain model fitted
values for the input parameters that are self-consistently determined;
they are the ionizing star characteristics, gas chemical abundances,
density, structure, and distance.

One of the main differences of this procedure when compared to
previous model calculations in the literature, is that we use the
distance as a fitting parameter for the model. This is possible
because we now have a way of producing projected images from the 3D
model results and can therefore compare them directly with observed
ones as well as the observed total fluxes. In other words, we do not
use a fixed distance for our model calculations and vary only star and
gas parameters (luminosity and temperature of the star, abundances and
densities of the gas). The other important advantage of using the 3-D
structure for the gas is the possibility of eliminating the need for a
``filling factor''. This has major implications on the model
parameters that can be determined, especially on the distance.

Since we use a 3-D structure that is consistent with observed images,
position-velocity diagrams, diagnostic ratios (such as density maps)
we are not making any assumptions about filling factors, ionized masses
or physical sizes. All these parameters are determined self-consistently
in the model.

The uniqueness of the solution obtained by this procedure can be
argued of course. It is immediately clear that within the observational
uncertainties there are an infinite number of solutions that can fit
the observations. In other words, the observations determine the
quality of the final parameters. In fact, if we estimate the goodness
of fit of our model fit by quadratically summing all uncertainties and
dividing by the number of observables minus the number of degrees of
freedom (input parameters to the model), we get an error of about 20\%
in the case of Mz1. This is the uncertainty adopted for our
results. The precise determination of fitting errors is extremely
complex and given the nature of the 3-D code neither practical nor
useful.

For the above reasons our distance determinations are fundamentally
different from, and much more precise than classically found
distances.

\begin{acknowledgements}
      Part of this work was supported by FAPESP grant 00/03126-5. We
      acknowledge the support of NOAO's Science Fund to
      HM. Nick Suntzeff's request for a ``back-of-the-envelope``
      explanation of our distance derivation helped us realize the
      importance of this method for nebular (nebulous?) research.
\end{acknowledgements}

\end{document}